\documentstyle[prl,aps,twocolumn]{revtex}
\begin{document}
\voffset 0.5in
\draft
\wideabs{
\title{Nonuniqueness of the Potentials of Spin-Density-Functional Theory}
\author{Klaus Capelle}
\address{Instituto de Qu\'{\i}mica de S\~ao Carlos,
Universidade de S\~ao Paulo, Caixa Postal 369, S\~ao Carlos, 13560-970 SP,
Brazil\\
and\\
Department of Physics and Astronomy, University of Missouri-Columbia,
Columbia, Missouri 65211, USA}
\author{G. Vignale}
\address{
Department of Physics and Astronomy, University of Missouri-Columbia,
Columbia, Missouri 65211, USA}
\date{\today}
\maketitle
\begin{abstract}
It is shown that, contrary to widely held beliefs, the potentials of 
spin-density-functional theory (SDFT) are not unique functionals of the 
spin densities. Explicit examples of distinct sets of 
potentials with the same ground-state densities are constructed, and general 
arguments that uniqueness should not occur in SDFT and other generalized
density-functional theories are given. 
As a consequence, various types of applications of SDFT require significant 
corrections or modifications.
\end{abstract}
}
\newcommand{\be}{\begin{equation}}
\newcommand{\ee}{\end{equation}}
\newcommand{\bea}{\begin{eqnarray}}
\newcommand{\eea}{\end{eqnarray}}
\newcommand{\bi}{\bibitem}

\newcommand{\ep}{\epsilon}
\newcommand{\s}{\sigma}
\newcommand{\p}{{\bf \pi}}
\newcommand{\r}{({\bf r})}
\newcommand{\rp}{({\bf r'})}
\newcommand{\rpp}{({\bf r''})}

\newcommand{\ua}{\uparrow}
\newcommand{\da}{\downarrow}
\newcommand{\la}{\langle}
\newcommand{\ra}{\rangle}
\newcommand{\dg}{\dagger}

The Hohenberg-Kohn (HK) theorem \cite{hk}, which guarantees that the 
ground-state single-particle density on its own is sufficient to uniquely 
determine all observables of a many-body system, is one of the 
most remarkable theorems of quantum mechanics.
It is also at the heart of density-functional theory (DFT), one of the most 
popular many-body methods \cite{dftbook,parryang}.
In the present paper a fundamental problem in the extension of this
theorem to spin-polarized situations is analysed, and consequences for
practical applications of SDFT are pointed out.

In the case of the original formulation of DFT, in which the basic
variable is the particle-density $n\r$, the HK theorem can be cast in
the form of two logically independent one-to-one maps \cite{dftbook}. 
Quantum mechanics guarantees that, for given particle-particle interaction
$\hat{U}$ and particle number $N$,
the potential in which the particles move determines the ground-state
many-body wave function (via solution of Schr\"odinger's equation), which in 
turn determines the ground-state density (by simple integration).
The essence of the original Hohenberg-Kohn theorem for nondegenerate ground 
states is that both of these maps are invertible: the ground-state density 
$n\r$ uniquely determines the ground-state wave function
$ \Psi({\bf r}_1,\ldots {\bf r}_N)$, which in turn determines, 
up to an additive constant, the potential \cite{hk,dftbook,parryang,levy},
\be
v\r \stackrel{(1)}{\Longleftrightarrow} \Psi({\bf r}_1,\ldots {\bf r}_N)
\stackrel{(2)}{\Longleftrightarrow} n\r.
\label{maps}
\ee
Map two implies that every ground-state observable is a unique functional of 
the 
density, whereas the combined map allows the much stronger statement that the 
potential (and, given the interaction, thus {\em every} observable) is a 
unique density functional.
These abstract maps have found extremely important practical 
applications in the form of the Kohn-Sham (KS) formulation of DFT \cite{ks},
which is used for almost all band-structure calculations in solid-state
physics, and a rapidly increasing number of electronic-structure calculations
in quantum chemistry.
Many of these applications, however, do not employ the original formulation of 
DFT, but spin-density-functional theory \cite{vbarthedin}, 
in which the fundamental variables are the spin-resolved particle densities
$n_\s\r$. In SDFT the map from spin densities to wave functions is easily 
established, but that from wave functions to potentials could, in spite of
considerable effort \cite{dftbook,vbarthedin,rajagopal} not be
proven and remains an, albeit popular, conjecture. 
In the early days of SDFT von Barth and Hedin \cite{vbarthedin} already pointed
out that the uniqueness of the spin-dependent potentials is not
guaranteed, and explicitly constructed two potentials which, when used
in a one-body Hamiltonian, have common eigenstates. The question whether
a similar construction is possible in the many-body case, however, remained 
open, and even in the one-body case it was objected that these
common eigenstates were not necessarily common ground states
\cite{dftbook,rajagopal}.

In this paper the general problem of uniqueness of the SDFT potentials is 
solved by (i) giving 
a general characterization (not limited to SDFT) of possible sources of 
nonuniqueness in DFT,
(ii) explicitly constructing different SDFT potentials with a common ground 
state, and (iii) pointing out the physical reason behind nonuniqueness.

To address point (i), let $v'=v+\Delta v$ and 
${\bf B}'={\bf B}+\Delta {\bf B}$ be electrostatic and magnetic
fields that are supposed to give the same (many-body) ground state $\Psi$ 
as $v$ and ${\bf B}$. $\Delta v$ and $\Delta {\bf B}$ then necessarily satisfy
\be
\int d^3r\, \left[\hat{n}\r \Delta v\r - \hat{\bf m}\r \Delta{\bf B}\r \right]
\Psi = \Delta E \Psi.
\label{condition}
\ee

A necessary condition for nonuniqueness is thus that, given $\Psi$, one
can find a linear combination of the density operators ($\hat{n}$ and
$\hat{m}$ in the case of SDFT) of which $\Psi$ is an eigenfunction.
This is automatically the case if a linear combination exists that is a
constant of motion, since constants of motion commute with the
Hamiltonian and thus have the same eigenfunctions (assumed nondegenerate
for convenience). If the Hamiltonian expressed in terms of $v$ and 
${\bf B}$ has a gap between its ground state and its first excited state,
one can always make sure that $\Psi$ remains ground state
of the new Hamiltonian, containing $v'$ and ${\bf B}'$, by making the 
coefficients of the linear combination, $\Delta v$ and
$\Delta {\bf B}$, sufficiently small.

From any extensive constant of motion (i.e., one which is a linear combination 
of the density operators) one can thus systematically construct families
of potentials with the same ground state. As an example consider
a system with an energy gap, for which the total magnetization 
$\hat{M}_z =\int d^3r\, \hat{m}_z\r $ is a constant
of motion. The choice $\Delta v=0$, $\Delta {\bf B} = \bar{B}{\bf u}_z 
=const$, where ${\bf u}_z$ is a unit vector in the z-direction,
then is clearly a suitable linear combination, as long as $\bar{B}$ is not
large enough to induce level crossings \cite{footnote0}.
The presence of a gap is crucially important, since it guarantees that for
sufficiently small values of $\bar{B}$ such level crossings do not
take place, so that $\Psi$ remains the ground state.

It is important to emphasize at this point that the exchange-correlation (xc)
functionals are still guaranteed to be unique density functionals by the 
{\em second} part of the HK theorem. Since the Hartree potential, too, is
manifestly a unique density functional, the nonuniqueness can only arise 
from the external potential contribution to the effective potentials. 
Practical consequences of this are pointed out towards the end of this
paper.

Examples for nonuniqueness obtained from constants of motion are 
below referred to as {\em systematic nonuniqueness}. Apart from these, 
Eq. (\ref{condition}) may, under suitable circumstances, also permit solutions 
not systematically associated with conserved quantities. 
To construct an example for such {\em accidental nonuniqueness} first recall 
how SDFT describes a spin-polarized system. After choosing the
xc functional and specifying the corresponding
spin-dependent Kohn-Sham potential $v_{s,\s}\r$,
one solves selfconsistently the spin-dependent Kohn-Sham equation,
\be
\left[-\frac{\hbar^2}{2m}\nabla^2 + v_{s,\s}\r\right]\phi_{n\s}\r=
\ep_{n\s} \phi_{n\s}\r,
\label{ks1}
\ee
and calculates the spin densities from
\be
n_\s\r = \sum_n^{N_\s} \phi_{n\s}^*\r \phi_{n\s}\r,
\label{density}
\ee
where $N_\s$, the number of particles with each spin, is determined by
distributing the $N$ particles over the $N$ lowest eigenfunctions $\phi_{n\s}$, 
such that the total energy is minimized.
If the number of filled spin up levels is equal to that of filled spin 
down levels, the system is unpolarized ($N_\ua =N_\da$), while it aquires
a finite spin polarization if they are different.
The maximum possible spin polarization is realized if all
particles have the same spin (say spin up), so that
$n_\ua = n$ and $n_\da=0$. 

In the case of a fully polarized system (e.g., saturated ferromagnetism), 
the N lowest eigenvalues of
the spin up Kohn-Sham equation are all lower than the lowest eigenvalue of
the spin down Kohn-Sham equation, so that no eigenfunction of the latter
is occupied. 
The full density is thus obtained from the selfconsistent solution of
\be
\left[-\frac{\hbar^2}{2m}\nabla^2 + v_{s,\ua}\r\right]\phi_{n\ua}\r=
\ep_{n\ua} \phi_{n\ua}\r
\label{ks2}
\ee
with
\be
n\r = n_\ua\r = \sum_n^N \phi_{n\ua}^*\r \phi_{n\ua}\r.
\ee
In this case the spin up Kohn-Sham potential of SDFT already 
determines the full particle density $n\r$.
The spin down potential, on the other hand, is a completely arbitrary
function of ${\bf r}$, as long as its lowest eigenvalue 
$\epsilon_{0\da}$ is higher than the $N$'th
eigenvalue of the corresponding spin up equation, $\epsilon_{N\ua}$
(which is precisely the condition for the system to remain fully polarized)
\cite{footnote1}.
The set of spin densities $\{n_\ua=n,n_\da=0\}$ thus does not uniquely
determine the set of potentials $\{v_{s\ua},v_{s\da}\}$, because
there trivially is an infinite number of
possible spin down potentials. Since spin densities and wave functions are 
still in one-to-one correspondence, this implies that the potentials of 
SDFT are not determined completely by either the KS Slater determinant $\Phi$
or the full many-body ground state $\Psi$.

In the more general case of a partially polarized system the above simple
argument for nonuniqueness breaks down, but it is not hard to show that 
partial polarization does not restore uniqueness.
Consider a given representation of the ground state KS Slater determinant
in terms of single-particle orbitals $\phi_n$. In ordinary DFT, based on the
density only, each of these orbitals determines the corresponding KS potential
(up to the irrelevant additive constant $\ep_n$ ) according to
\be
v_s\r = \ep_n - \frac{\hat{t}\phi_n\r}{\phi_n\r}, 
\label{vsolve1}
\ee
where $\hat{t}$ is the single-particle kinetic energy operator.
Eq. (\ref{vsolve1}) is simply the Kohn-Sham equation solved for 
$v_s\r$.
If one now tries to repeat this reasoning in the spin-dependent case, one 
must solve the Kohn-Sham equation of SDFT for the external potentials
$v_{s,\ua}\r$ and $v_{\s,\da}\r$ in terms of the single-particle orbitals
$\phi_{n\ua}\r$ and $\phi_{n\da}\r$,
\be
v_{s,\s}\r = \ep_{n\s} - \frac{\hat{t}\phi_{n\s}\r}{\phi_{n\s}\r}.
\label{vsolve2}
\ee
Here we restricted us to collinear spin configurations, so that the
magnetic field can be expressed in terms of the spin-dependent potentials
as $B_z\r = [v_{s,\da}\r - v_{s,\ua}\r]/\mu$, where $\mu = q\hbar/(2mc)$
is the Bohr magneton.

Again, any of the single-particle orbitals determines the corresponding 
potential up to a constant, which remains unspecified. But, unlike in the
previous case {\em one can choose only one constant freely by adjusting
the zero of energy}. That is, given $\phi_{n\ua}\r$, say, one can
determine $v_{s,\ua}\r$ up to an irrelevant constant, but then the zero 
of energy has been fixed, and $\phi_{n\da}\r$ determines
$v_{s,\da}\r$ only up to a (now physically relevant) constant, rigidly
shifting spin up levels with respect to spin down levels.
Consequently, the KS Slater determinant does not completely determine
the spin-dependent potentials. By appealing to the second part of 
the HK theorem one readily concludes that the same holds for the many-body
wave function $\Psi$.

The only caveat to the above construction is that for the original KS 
determinant
to remain the ground state after adding the free constant to $v_{s,\da}$, 
this shift must not induce the presence of new (previously unoccupied) 
single-particle orbitals in the Slater determinant.
As above, this is guaranteed if the KS N-particle ground state is separated 
from the first excited state by a gap, and the shift of $v_{s,\da}$ is 
sufficiently small compared to this gap.

Adding a constant term, $\bar{v}$, to $v_{s,\da}\r$ amounts to adding $\bar{v}$ 
to the electrostatic potential $v_s\r$ and $\bar{v}/\mu$ to the
magnetic field $B_z\r$. The nonuniqueness associated with the fact that there 
is only one zero of energy, but two constants to be determined in  
Eq. (\ref{vsolve2}), has thus brought us back to a particular example
of systematic nonuniqueness, in which the linear combination in 
Eq. (\ref{condition}) is
given by $\Delta v = \bar{v}$ and $\Delta {\bf B} = (\bar{v}/\mu){\bf u}_z$.  
In general we expect that systematic nonuniqueness is the only generic source
of nonuniqueness, while accidental nonuniqueness requires certain 
`pathological' features, such as complete spin polarization.

Reflecting on the previous examples, what all cases of nonuniqueness, 
accidental or systematic, have in common is that the wave functions do not 
change if certain changes 
are made to the potentials. This implies a degree of 'rigidity' or 
'incompressibility' of the former with respect to the latter.
The relation to incompressibility is, in fact, more than a mere analogy:
an essential feature of all nontrivial examples 
of nonuniqueness (i.e., all except the 
additive constant one can choose in the electrostatic potential)
is a gap between ground and first excited state, as a consequence of which
sufficiently small changes to the potentials cannot bring down an excited 
state to have lower energy than the ground state.
The existence of a gap between ground and first excited state, on the
other hand, is tantamount to incompressibility in the usual 
(quantum liquid) sense of the word.
Such incompressibility thus constitutes a
necessary (but not sufficient) condition for nontrivial nonuniqueness.

In the remainder of this paper we look at several applications of SDFT in 
order to determine what changes become necessary in view of the above results.
First of all, in the standard Kohn-Sham formulation of SDFT the 
nonuniqueness of $v_{s,\s}$ is not critical, since it arises from the 
external potentials (which are known from the beginning), whereas the xc 
potentials remain unique.
Fields in which significant changes are required if the first map is no 
longer available, include (i) the calculation of excited states, 
(ii) the determination of exact KS potentials from numerically exact densities,
(iii) perturbative strategies to construct approximate density functionals,
(iv) the optimized-effective potential method of SDFT, and 
(v) several recent generalizations of DFT.

(i) Let us consider excited states first.
A standard argument \cite{dftbook} to justify the calculation of excited
states in DFT (albeit not by direct identification of the Kohn-Sham
eigenvalues with true excitation energies) is that these, too, are 
functionals of the ground-state density; afterall, according to the combined 
map $v[\Psi[n]]$ this density
determines the full potential, which, via Schr\"odinger's
equation, determines all eigenstates, ground state as well as excited ones.
Clearly, this argument is only correct in the original formulation
of DFT, in which the combined map is available. The above discussion 
shows that it cannot be used to justify attempts to employ SDFT 
for the calculation of excited states \cite{footnote3}.

(ii) Recently several methods for constructing accurate KS potentials, 
using as input highly precise numerical densities (obtained from 
Monte Carlo or configuration interaction calculations), have been proposed
\cite{parr1,parr2,goerling,arya,robert,march,handy}. 
Uniqueness of the resulting 
KS potentials is taken for granted in these prescriptions.
Again, this holds only in the original formulation of DFT, but not in the 
(much more widely used) SDFT. The generalization of these methods to determine 
the KS potentials of SDFT must thus be reconsidered. Even the xc potentials, 
which are unique, are not completely free of this problem: in an exact 
calculation 
the nonuniqueness of the KS potentials must of course cancel in the final 
result, but this is not guaranteed to be the case in an actual (necessarily 
approximate) implementation, e.g., using finite basis sets. Special
care must thus be taken in such calculations to ensure that the xc potential
is not contaminated by nonuniqueness.

(iii) A useful method for obtaining LDA-type approximations for the 
xc functional is based on perturbative diagrammatic calculations of 
the xc energy of the electron gas \cite{dftbook,scdftlett}.
The results of such calculations are necessarily obtained in terms of the 
potentials acting on the unperturbed system. As long as the first map of
the HK theorem holds, they are thus automatically also functionals of the
densities. Obviously this ceases to be true if that map is no longer 
available. It follows from the above that the SDFT counterpart of this 
procedure works only because the electron gas does not have a gap between
its ground and first excited state. Attempts to apply this procedure to
systems with a gap (such as the superconducting electron gas considered
in Ref. \cite{scdftlett}) thus face unanticipated and fundamental problems 
in their generalization to the spin-resolved case.

(iv) In the optimized-effective potential (OEP) method one minimizes an energy
functional that is an explicit functional of the single-particle orbitals
and only an implicit functional of the densities. In order to calculate the 
functional derivative with respect to the densities, one then employs the
functional chain rule, to write \cite{oep1,oep2} 
\bea
v_{xc}[n]\r=
\frac{\delta E_{xc}[\{\phi_n\}]}{\delta n\r}=
\nonumber \\
\int dr' \int dr'' \sum_n
\frac{\delta E_{xc}[\{\phi_n\}]}{\delta \phi_n\rp}
\frac{\delta \phi_n\rp}{\delta v\rpp}
\frac{\delta v\rpp}{\delta n\r} +c.c.
\label{oepchain}
\eea
The first functional derivative on the right-hand side can be calculated 
explicitly, the second is easily found from perturbation theory, and the last 
is usually identified with an inverse response function. In SDFT 
the potentials are not unique, so that this inverse does not exist for all 
densities, and hence the solution to the OEP integral equation following from 
Eq. (\ref{oepchain}) is not guaranteed to be the correct $v_{xc}$.
The problem here arises from using the KS potential as an intermediate
step in performing the functional derivative of the orbital functional, the
xc potential itself is of course still unique. This intermediate step
can be avoided by directly using $\delta \phi_n/\delta n$ on
the right-hand side, but the resulting equation will be more complicated to 
solve than the standard OEP equation.

(v) Finally, consider generalized density-functional theories.
The fact that even in the oldest and most widely used of all
generalizations of DFT, namely SDFT, uniqueness of the potentials
is not guaranteed, casts serious doubts on uniqueness claims in
other, more recent, generalizations of DFT. It can be suspected
that in, e.g., relativistic DFT, time-dependent DFT, polarization-dependent
DFT, DFT for superconductors, etc, similar
problems to those discussed here in the context of SDFT are lurking.
Indeed, in current-DFT \cite{vr2} this is known to be the case, and examples 
of both systematic and accidental nonuniqueness can be found \cite{gvunpubl}.

In summary, it has been shown, both by explicit examples and by general 
considerations, that, unlike in ordinary DFT, in SDFT and other generalized 
DFTs the effective and the external potentials are not uniquely determined 
by the spin densities
alone. Nonuniqueness can arise accidentally, via special features of the
ground state, and systematically, via extensive constants of motion.
In both cases it depends on the presence of a gap in the spectrum of the
system, and on the resulting incompressibility.
As a consequence many previous applications of SDFT and other 
generalized DFTs in condensed 
matter physics and quantum chemistry must be critically reexamined. 

{\bf Acknowledgments}
KC thanks L. N. Oliveira, M. L\"uders, and E. K. U. Gross for 
valuable discussions, the Physics Department at the University of Columbia,
Missouri, for generous hospitality, and the FAPESP for financial support.
GV acknowledges support from NSF Grant DMR 9706788.

\end{document}